\documentclass[runningheads]{llncs}
\usepackage[T1]{fontenc}
\usepackage{graphicx,verbatim}
\usepackage{color}

\usepackage[symbol]{footmisc}

\usepackage[numbers]{natbib}
\usepackage{hyperref}
\usepackage{mathtools}
\usepackage{graphicx}
\usepackage{multirow}
\usepackage{subcaption}
\usepackage{booktabs}

\usepackage{tabularx}
\usepackage{makecell}          
\newcommand{\valci}[3]{\makecell{#1\\[-1pt]\tiny (#2–#3)}} 

\usepackage{layout}

\begin{document}

\title{Domain-Agnostic Stroke Lesion Segmentation Using Physics-Constrained Synthetic Data}

\author{Liam Chalcroft\inst{1}\orcidID{0000-0003-3363-6454} \and Jenny Crinion\inst{2}\orcidID{0000-0001-8080-6562} \and Cathy J. Price\inst{1}\orcidID{0000-0001-7448-4835} \and John Ashburner\inst{1}\orcidID{0000-0001-7605-2518}}


\authorrunning{L. Chalcroft et al.}
\titlerunning{Physics-Constrained Synthetic Data}

\institute{
Department of Imaging Neuroscience, University College London
\and
Institute of Cognitive Neuroscience, University College London
\\ \email{l.chalcroft@cs.ucl.ac.uk}
}


\maketitle             

\begin{abstract}
Segmenting stroke lesions in MRI is challenging due to diverse acquisition protocols that limit model generalisability. In this work, we introduce two physics-constrained approaches to generate synthetic quantitative MRI (qMRI) images that improve segmentation robustness across heterogeneous domains. Our first method, \texttt{qATLAS}, trains a neural network to estimate qMRI maps from standard MPRAGE images, enabling the simulation of varied MRI sequences with realistic tissue contrasts. The second method, \texttt{qSynth}, synthesises qMRI maps directly from tissue labels using label-conditioned Gaussian mixture models, ensuring physical plausibility. Extensive experiments on multiple out-of-domain datasets show that both methods outperform a baseline UNet, with \texttt{qSynth} notably surpassing previous synthetic data approaches. These results highlight the promise of integrating MRI physics into synthetic data generation for robust, generalisable stroke lesion segmentation. Code is available at \url{https://github.com/liamchalcroft/qsynth}
\end{abstract}


\section{Introduction} \label{intro}

Segmenting brain pathologies in MRI is vital for clinical and research applications, yet remains challenging due to the variability of acquisition protocols across hospitals. Recent work shows that even for diffusion-weighted MRI - arguably the modality of choice in acute stroke - dedicated CNNs can equal human raters when trained on large, carefully curated cohorts \cite{Liu2021}. While public datasets achieve high performance using standardised sequences (e.g., T1w, T2w, FLAIR) with similar parameters, clinical data rarely conforms to these ideal conditions \cite{delarosa2024robustensemblealgorithmischemic, Brzus2023}. Existing domain adaptation methods either require prior target domain knowledge or many unlabelled images \cite{Dorent2023, wang2021tent}, and synthetic approaches like SynthSeg \cite{Billot2023} can produce unrealistic contrasts for heterogeneous stroke lesions \cite{Billot2021,chalcroft2024syntheticdatarobuststroke}.

To address these limitations, we propose generating synthetic images using quantitative MRI (qMRI) parameters and physics-based forward models to ensure physical plausibility. qMRI provides voxel-level tissue properties - proton density (PD), longitudinal relaxation rate ($R_1$), effective transverse relaxation rate ($R_2^*$), and magnetisation transfer (MT) - that enable the simulation of diverse MRI sequences while preserving tissue characteristics. Since acquiring full qMRI data is time-consuming, recent deep learning methods estimate qMRI maps from standard sequences \cite{borges2021acquisitioninvariant,varadarajan2021unsupervised,Borges2023}, making large-scale synthetic data generation feasible.

We introduce two methods for domain-agnostic stroke lesion segmentation. The first, \texttt{qATLAS}, trains a qMRI estimation model on MPRAGE images to augment the ATLAS dataset \cite{Liew2022} with simulated sequences. The second, \texttt{qSynth}, extends previous synthetic approaches \cite{Billot2021,chalcroft2024syntheticdatarobuststroke} by sampling qMRI maps from intensity priors derived from real data, ensuring greater physical realism. By embedding MRI physics into data synthesis, our framework bridges the gap between synthetic and clinical data, thereby improving segmentation robustness across diverse imaging domains.

\section{Methods}

We propose two methods, \texttt{qATLAS} and \texttt{qSynth}, for domain-agnostic stroke lesion segmentation. Both leverage qMRI parameter maps to generate diverse, physics-constrained training data, enhancing robustness and generalisability.

\subsection{qATLAS: Estimating qMRI from MPRAGE}

\texttt{qATLAS} fits an nnU-Net \cite{Isensee2020} that predicts four qMRI maps (PD, $R_1$, $R_2^{*}$, MT) from a single MPRAGE\footnote{Estimating $R_2^{*}$ from one T1 is ill-posed; errors mainly affect simulated T2w/FLAIR.}. Training used 51 subjects (22 healthy, 29 stroke; 80/20 split) with 3D-EPI ground-truth maps from hMRI \cite{Tabelow2019}. Thus any routine MPRAGE can be converted to qMRI and fed to our physics-based simulator.

Diverse MPRAGE images were simulated using NiTorch with parameters: $T_R \sim \mathcal{U}(1.9,2.5)$ s, $T_I \sim \mathcal{U}(0.6,1.2)$ s, $T_E \sim \mathcal{U}(2,4)$ ms, $\alpha \sim \mathcal{U}(5^\circ,12^\circ)$, and $B_0 \sim \mathcal{U}(0.3,7)$ T.

\begin{figure}[hbt!]
    \centering
    \includegraphics[width=\textwidth]{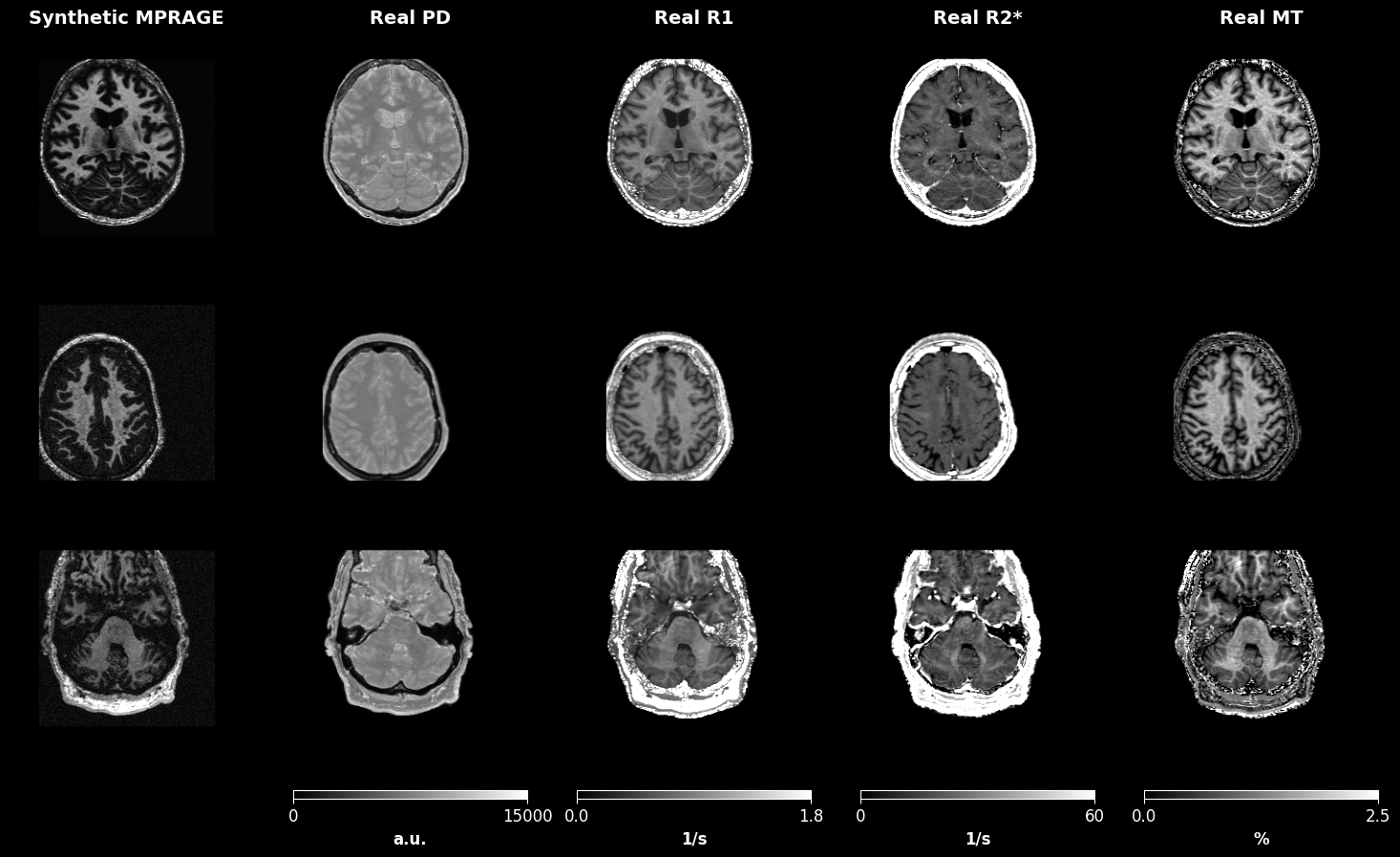}
    \caption{Examples of training data for qMRI parameter map prediction in the qATLAS method.}
    \label{fig:mprage-mpm-viz}
\end{figure}

\textbf{Data Augmentation:} We applied standard augmentations (elastic/affine deformations, bias field, Gibbs ringing, Rician noise, and random cropping to $192^3$ voxels) using MONAI \cite{cardoso2022monai}. Figure \ref{fig:mprage-mpm-viz} illustrates examples of the augmented training data.

\textbf{Model Architecture and Training:} Our U-Net comprises five encoder stages (channels: 24, 48, 96, 192, 384), each with two residual units, using GELU activations \cite{hendrycks2023gaussian}, instance normalisation \cite{ulyanov2017instance}, linear upsampling, and 0.1 dropout. The network outputs four channels corresponding to PD, $R_1$, $R_2^*$, and MT. To enforce positivity, PD, $R_1$, and $R_2^*$ are computed as the exponential of their raw outputs, while MT is computed as 100 times the sigmoid of its raw output.

Training was performed over 200,000 iterations (batch size 1) using AdamW (lr=$10^{-4}$, $\beta_1=0.9$, $\beta_2=0.999$, weight decay=0.01). An initial 20,000 iterations used L2 loss; thereafter, we employed a combined loss $L = \text{L1}(y,\hat{y}) + \text{L2}(y,\hat{y}) + \text{L1}(\nabla y,\nabla \hat{y}) + \text{L2}(\nabla y,\nabla \hat{y})$ augmented by a perceptual LPIPS loss \cite{zhang2018unreasonable} (weight 0.1) using features from a pretrained Med3D encoder \cite{chen2019med3d}.

\begin{figure}[hbt!]
    \centering
    \includegraphics[width=0.8\textwidth]{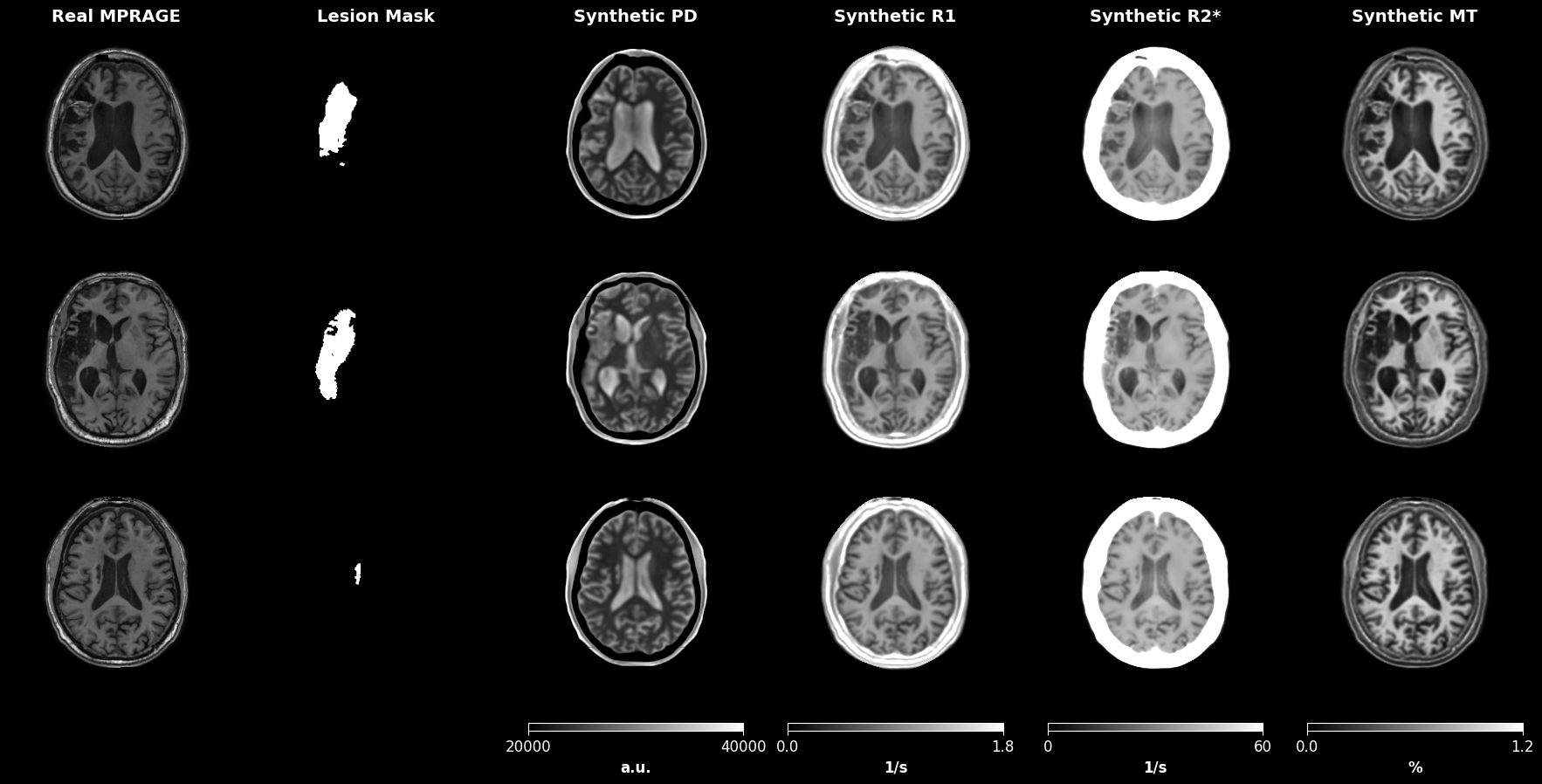}
    \caption{Examples of predicted qMRI parameter maps from input MPRAGE images of the ATLAS dataset in the \texttt{qATLAS} method.}
    \label{fig:atlas-mpm-viz}
\end{figure}

Figure \ref{fig:atlas-mpm-viz} displays examples of predicted qMRI maps from input MPRAGE images in the ATLAS dataset \cite{Liew2022}.

\textbf{Simulation of MRI Sequences:} We used the estimated qMRI maps from the ATLAS dataset to simulate diverse MRI sequences via a physics-based generative model (see Section \ref{sec:methods-physics}), thereby creating the \texttt{qATLAS} dataset for segmentation training.

\subsection{qSynth: Synthesising qMRI Maps from Tissue Labels}

The \textbf{qSynth} method generates synthetic qMRI maps directly from segmentation labels using label-conditioned Gaussian Mixture Models (GMMs). We first define prior distributions for each qMRI parameter (PD, $R_1$, $R_2^*$, MT) for different tissue types (gray matter, white matter, cerebrospinal fluid, and lesions). These priors are estimated from the population used in \texttt{qATLAS} (for anatomy or pathology where real data is not available, it is feasible that parameters could be estimated based on values reported in literature). For each voxel in a segmentation label, the corresponding qMRI parameters are sampled from the appropriate prior, thereby producing a synthetic qMRI map that reflects realistic tissue properties. Healthy tissue maps are obtained using the Multibrain SPM toolbox \cite{Brudfors_2020}, and lesion masks are generated by overlaying random binary lesion maps onto these healthy tissue maps (see \cite{chalcroft2024syntheticdatarobuststroke} for further details).

\textbf{Simulation of MRI Sequences:} The synthetic qMRI maps are then used with our physics-based generative model (Section \ref{sec:methods-physics}) to simulate diverse MRI sequences, resulting in the \textbf{qSynth} dataset for training an alternative segmentation model.

\subsection{Physics-Based Generative Model} \label{sec:methods-physics}

Both \texttt{qATLAS} and \texttt{qSynth} use a physics-based generative model to simulate realistic MRI images from qMRI parameter maps via standard signal equations. We simulate several common MRI sequences, including Fast Spin-Echo (FSE), Gradient-Echo (GRE), Fluid-Attenuated Inversion Recovery (FLAIR), and Magnetisation-Prepared Rapid Gradient Echo (MPRAGE).

For example, the FSE signal is computed as:
\[
S_{\text{FSE}} = B_1\,\text{PD}\,\Bigl(1 - e^{-R_1T_R}\Bigr)e^{-R_2T_E},
\]
where \(B_1\) is the receive field strength, PD is the proton density, \(R_1\) and \(R_2\ (\approx R2^*)\) are the relaxation rates, and \(T_R\) and \(T_E\) are the repetition and echo times, respectively. Similar equations are used for GRE, FLAIR, and MPRAGE sequences.

Acquisition parameters (e.g., \(T_R\), \(T_E\), \(T_I\), \(T_X\), \(T_D\), and \(\alpha\)) were sampled from physically plausible distributions; please refer to our code repository\footnote{\url{https://github.com/liamchalcroft/qsynth}} for the complete sampling details.

To mimic realistic MRI data, we add Rician noise by perturbing the simulated signal \(S_{\text{MRI}}\) with independent Gaussian noise on the real and imaginary components:
\[
S_{\text{noisy}} = \sqrt{(S_{\text{MRI}} + n_r)^2 + n_i^2},\quad n_r,n_i\sim\mathcal{N}(0,\sigma^2).
\]

All simulations and noise additions were implemented using the NiTorch library\footnote{\url{https://github.com/balbasty/nitorch}}. Additional augmentations - such as elastic deformations, bias field variations, axis flips, Gaussian noise, low-resolution reslicing, and random cropping to \(192\times192\times192\) voxels - were applied using MONAI \cite{cardoso2022monai}. 


\subsection{Segmentation Model Training}

We trained nnUNet-based segmentation models \cite{Isensee2020} on four settings: (i) real MPRAGE images from the ATLAS dataset (\texttt{baseline}), (ii) the \texttt{qATLAS} dataset (from MPRAGE-derived qMRI maps), (iii) a synthetic data model using the public \texttt{Synth} method \cite{chalcroft2024syntheticdatarobuststroke}, and (iv) the \texttt{qSynth} method. The \texttt{qSynth} method differs from \texttt{Synth} in that \texttt{Synth} directly samples the random tissue intensities, while \texttt{qSynth} uses intensity priors to first synthesise qMRI parameter maps, which may then be used to generate realistic tissue intensities via random variations of real forward models. For \texttt{Synth} and \texttt{qSynth}, we additionally evaluated models trained on a mixture of synthetic and real ATLAS images.

\textbf{Model Architecture and Training Details:}  
All models use PReLU activations \cite{he2015delving} with one residual unit per block. The \texttt{qATLAS} model performs binary segmentation (background vs. stroke lesion), while the \texttt{qSynth} model predicts additional healthy tissue classes (gray matter, white matter, GM/WM partial volume, and cerebrospinal fluid). Baseline and \texttt{qATLAS} models were trained on the ATLAS dataset (N=419/105/131 train/validation/test), whereas \texttt{Synth} and \texttt{qSynth} models used ATLAS lesion labels combined with healthy tissue labels from OASIS-3 (N=2579/100 train/validation) as detailed in \cite{chalcroft2024syntheticdatarobuststroke}.

All models were optimised using a combined Dice and cross-entropy loss with the AdamW optimiser \cite{loshchilov2019decoupled} (lr=$10^{-4}$, $\beta_1=0.9$, $\beta_2=0.999$, weight decay=0.01) and a learning rate scheduler $\eta_n=\eta_0\,(1-\frac{n}{N})^{0.9}$. Training was run for 700,000 iterations with a batch size of 1.

\section{Experiments}

We evaluated segmentation models on four datasets: ATLAS, comprising 131 subjects (isotropic MPRAGE) \cite{Liew2022} for in-domain evaluation; ARC, with 229 subjects (T1w, T2w, FLAIR) \cite{Gibson2024TheRepository,Johnson2024ProgressiveStroke} for out-of-domain testing; PLORAS, consisting of 406 subjects (106 T2w, 300 FLAIR) from various UK hospitals; and ISLES 2015, including 28 subjects (T1w, T2w, FLAIR, DWI) \cite{Maier2017} with acute lesions.

All scans were resliced to 1 mm, histogram–normalised and $z$-scored, then segmented with a $192^{3}$ sliding window (50 \% overlap, Gaussian blending) and 8-flip TTA \cite{Wang_2019}. For multi-modal sets (ARC, ISLES15) we averaged per-voxel logits across contrasts before soft-max and post-processing (reported as 'Ensemble').  Dice and HD95 (95-percentile Hausdorff, mm) were computed on predictions/GT padded to $256^{3}$; only the \emph{lesion} channel is reported for \texttt{Synth}/\texttt{qSynth}.

\noindent\textbf{Dataset caveat.}  
\texttt{Synth}/\texttt{qSynth} are trained on a much larger multi-site cohort with multi-class labels, whereas \texttt{qATLAS} and the baseline use ATLAS only (binary masks). Thus absolute score differences combine physics, training-set size, label granularity and pre-stripped skulls in ISLES15; isolating each factor is left for future work.

\section{Results}

\begin{figure}[hbt!]
    \centering
    \begin{subfigure}[t]{\textwidth}
        \centering
        \includegraphics[width=\textwidth]{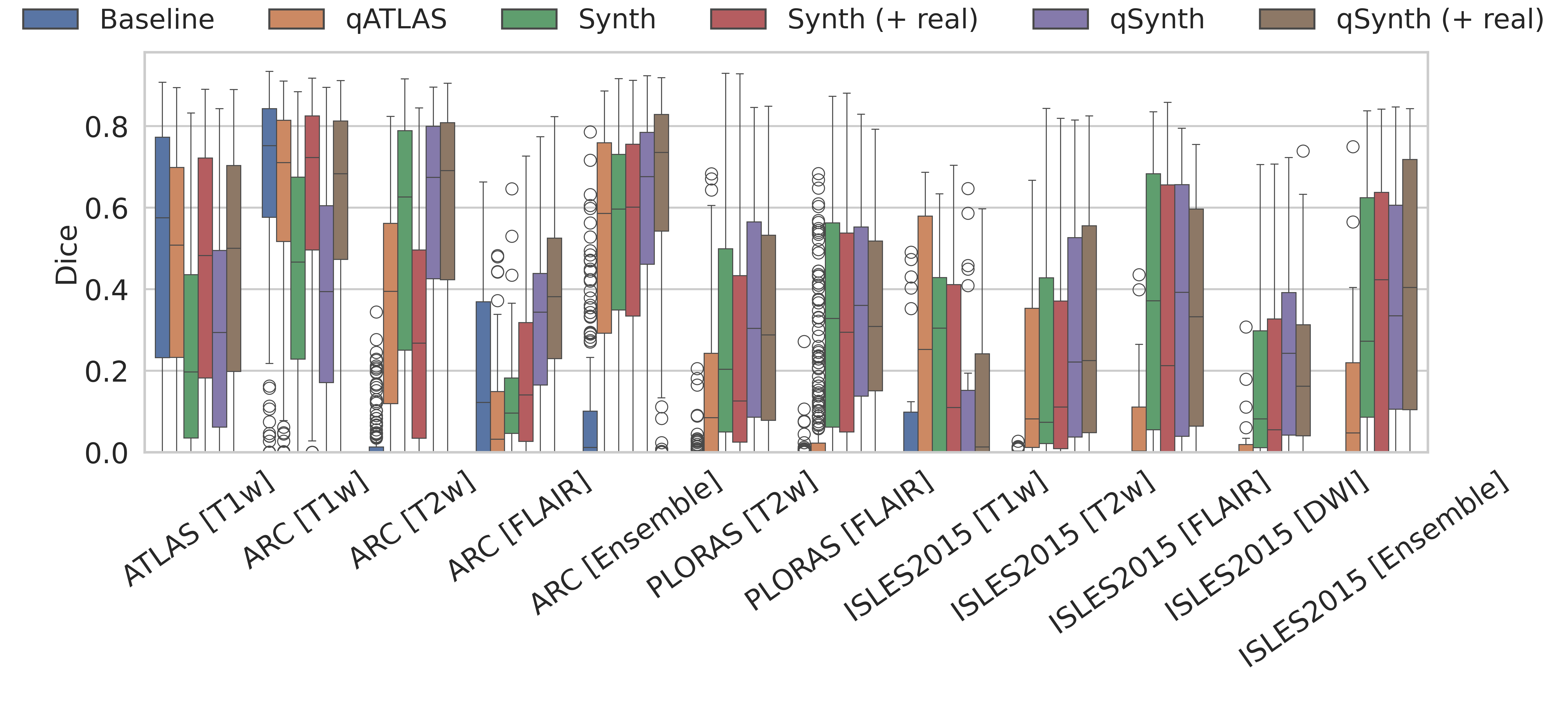}
    \end{subfigure}
    \caption{Dice metric performance for all datasets.}
    \label{fig:combined-box}
\end{figure}

\begin{table}[t]
\caption{Median \textbf{Dice} (top line) and 95 \% CI (tiny line). Best, second, and third best are \textbf{bold}, \underline{underlined}, and \textit{italic}.}
\centering
\scriptsize
\setlength{\tabcolsep}{3pt}
\begin{tabularx}{\textwidth}{@{}ll*{6}{c}@{}}
\toprule
Dataset & Modality &
Baseline & qATLAS & Synth & Synth\,+Real & qSynth & qSynth\,+Real\\
\midrule
\multirow{1}{*}{ATLAS}
 & T1w &
 \textbf{\valci{0.575}{0.522}{0.628}} &
 \underline{\valci{0.508}{0.460}{0.556}} &
 \valci{0.197}{0.155}{0.239} &
 \valci{0.482}{0.431}{0.534} &
 \valci{0.294}{0.250}{0.338} &
 \textit{\valci{0.501}{0.450}{0.551}} \\[2pt]
\midrule
\multirow{4}{*}{ARC}
 & T1w &
 \textbf{\valci{0.752}{0.713}{0.790}} &
 \textit{\valci{0.710}{0.673}{0.747}} &
 \valci{0.467}{0.430}{0.504} &
 \underline{\valci{0.723}{0.684}{0.762}} &
 \valci{0.394}{0.359}{0.429} &
 \valci{0.683}{0.643}{0.723} \\
 & T2w &
 \valci{0.000}{0.000}{0.007} &
 \valci{0.395}{0.363}{0.426} &
 \textit{\valci{0.626}{0.588}{0.665}} &
 \valci{0.268}{0.233}{0.302} &
 \underline{\valci{0.674}{0.640}{0.709}} &
 \textbf{\valci{0.691}{0.655}{0.726}} \\
 & FLAIR &
 \valci{0.122}{0.076}{0.169} &
 \valci{0.032}{0.005}{0.059} &
 \valci{0.096}{0.070}{0.123} &
 \textit{\valci{0.141}{0.097}{0.185}} &
 \underline{\valci{0.344}{0.300}{0.388}} &
 \textbf{\valci{0.382}{0.332}{0.431}} \\
 & Ensemble &
 \valci{0.012}{0.000}{0.032} &
 \valci{0.586}{0.549}{0.622} &
 \valci{0.597}{0.562}{0.631} &
 \textit{\valci{0.602}{0.566}{0.638}} &
 \underline{\valci{0.676}{0.642}{0.709}} &
 \textbf{\valci{0.735}{0.701}{0.770}} \\[2pt]
\midrule
\multirow{2}{*}{PLORAS}
 & T2w &
 \valci{0.000}{0.000}{0.006} &
 \valci{0.085}{0.049}{0.122} &
 \textit{\valci{0.204}{0.150}{0.258}} &
 \valci{0.126}{0.073}{0.179} &
 \textbf{\valci{0.304}{0.253}{0.356}} &
 \underline{\valci{0.288}{0.238}{0.338}} \\
 & FLAIR &
 \valci{0.000}{0.000}{0.002} &
 \valci{0.000}{0.000}{0.017} &
 \underline{\valci{0.328}{0.298}{0.358}} &
 \valci{0.294}{0.264}{0.325} &
 \textbf{\valci{0.361}{0.333}{0.388}} &
 \textit{\valci{0.309}{0.283}{0.335}} \\[2pt]
\midrule
\multirow{5}{*}{ISLES15}
 & T1w &
 \valci{0.000}{0.000}{0.064} &
 \underline{\valci{0.252}{0.142}{0.363}} &
 \textbf{\valci{0.304}{0.213}{0.396}} &
 \textit{\valci{0.110}{0.012}{0.208}} &
 \valci{0.002}{0.000}{0.078} &
 \valci{0.013}{0.000}{0.087} \\
 & T2w &
 \valci{0.000}{0.000}{0.002} &
 \valci{0.082}{0.000}{0.170} &
 \valci{0.074}{0.000}{0.184} &
 \textit{\valci{0.111}{0.007}{0.216}} &
 \underline{\valci{0.222}{0.114}{0.329}} &
 \textbf{\valci{0.225}{0.117}{0.333}} \\
 & FLAIR &
 \valci{0.000}{0.000}{0.000} &
 \valci{0.004}{0.000}{0.052} &
 \underline{\valci{0.372}{0.255}{0.489}} &
 \valci{0.212}{0.085}{0.340} &
 \textbf{\valci{0.392}{0.281}{0.504}} &
 \textit{\valci{0.332}{0.226}{0.439}} \\
 & DWI &
 \valci{0.000}{0.000}{0.000} &
 \valci{0.001}{0.000}{0.027} &
 \textit{\valci{0.082}{0.000}{0.168}} &
 \valci{0.056}{0.000}{0.144} &
 \textbf{\valci{0.243}{0.155}{0.331}} &
 \underline{\valci{0.162}{0.076}{0.249}} \\
 & Ensemble &
 \valci{0.000}{0.000}{0.000} &
 \valci{0.048}{0.000}{0.123} &
 \valci{0.272}{0.157}{0.388} &
 \textbf{\valci{0.423}{0.302}{0.545}} &
 \textit{\valci{0.335}{0.222}{0.448}} &
 \underline{\valci{0.404}{0.284}{0.524}} \\
\bottomrule
\end{tabularx}
\label{tab:dice}
\end{table}

\begin{table}[t]
\caption{95th-percentile Hausdorff distance \textbf{(HD95, mm)} - median on top, 95 \% CI beneath. Lowest, 2nd-lowest, and 3rd-lowest scores are \textbf{bold}, \underline{underlined}, and \textit{italic}.}
\centering
\scriptsize
\setlength{\tabcolsep}{3pt}
\begin{tabularx}{\textwidth}{@{}ll*{6}{c}@{}}
\toprule
Dataset & Modality &
Baseline & qATLAS & Synth & Synth\,+Real & qSynth & qSynth\,+Real\\
\midrule
\multirow{1}{*}{ATLAS}
 & T1w &
 \textbf{\valci{19.7}{10.2}{29.3}} &
 \textit{\valci{34.4}{26.5}{42.2}} &
 \valci{63.7}{59.6}{67.9} &
 \underline{\valci{22.6}{12.0}{33.3}} &
 \valci{51.3}{43.5}{59.2} &
 \valci{38.0}{28.2}{47.8} \\[2pt]
\midrule
\multirow{4}{*}{ARC}
 & T1w &
 \textbf{\valci{8.8}{2.5}{15.0}} &
 \textit{\valci{11.8}{5.6}{18.1}} &
 \valci{57.1}{52.9}{61.3} &
 \underline{\valci{11.0}{4.0}{18.0}} &
 \valci{31.0}{26.6}{35.5} &
 \valci{13.0}{6.8}{19.2} \\
 & T2w &
 \valci{74.4}{71.4}{77.4} &
 \valci{54.5}{50.7}{58.2} &
 \textit{\valci{46.1}{42.2}{50.0}} &
 \valci{64.0}{60.5}{67.4} &
 \textbf{\valci{39.0}{34.8}{43.2}} &
 \underline{\valci{43.0}{38.7}{47.2}} \\
 & FLAIR &
 \valci{58.5}{49.5}{67.4} &
 \valci{59.7}{50.7}{68.7} &
 \valci{57.6}{49.2}{65.9} &
 \textit{\valci{52.2}{43.3}{61.1}} &
 \underline{\valci{50.3}{41.3}{59.2}} &
 \textbf{\valci{44.0}{34.5}{53.5}} \\
 & Ensemble &
 \valci{48.2}{40.1}{56.3} &
 \textbf{\valci{16.4}{10.9}{22.0}} &
 \valci{44.0}{39.9}{48.2} &
 \underline{\valci{20.1}{14.8}{25.3}} &
 \valci{31.6}{27.2}{36.0} &
 \textit{\valci{20.7}{16.2}{25.2}} \\[2pt]
\midrule
\multirow{2}{*}{PLORAS}
 & T2w &
 \valci{70.9}{65.9}{76.0} &
 \textbf{\valci{59.6}{54.4}{64.8}} &
 \valci{67.5}{61.3}{73.7} &
 \valci{70.3}{64.9}{75.8} &
 \underline{\valci{60.4}{54.6}{66.3}} &
 \textit{\valci{60.7}{55.2}{66.3}} \\
 & FLAIR &
 \underline{\valci{66.8}{61.2}{72.4}} &
 \textbf{\valci{65.4}{58.4}{72.3}} &
 \textit{\valci{69.7}{66.8}{72.6}} &
 \valci{74.3}{71.4}{77.2} &
 \valci{73.3}{70.4}{76.2} &
 \valci{73.0}{70.2}{75.7} \\[2pt]
\midrule
\multirow{5}{*}{ISLES15}
 & T1w &
 \valci{69.8}{30.0}{109.7} &
 \textbf{\valci{51.5}{40.0}{63.0}} &
 \valci{58.6}{48.5}{68.6} &
 \underline{\valci{52.5}{29.9}{75.0}} &
 \textit{\valci{56.3}{19.6}{93.1}} &
 \valci{56.3}{17.8}{94.8} \\
 & T2w &
 \underline{\valci{63.1}{57.8}{68.4}} &
 \textbf{\valci{59.4}{43.3}{75.4}} &
 \valci{72.3}{65.7}{78.9} &
 \valci{75.0}{68.4}{81.7} &
 \textit{\valci{66.1}{58.3}{73.8}} &
 \valci{67.7}{59.8}{75.6} \\
 & FLAIR &
 \valci{71.8}{38.5}{105.2} &
 \textit{\valci{65.8}{51.2}{80.5}} &
 \textbf{\valci{56.1}{49.5}{62.7}} &
 \textbf{\valci{56.1}{38.7}{73.4}} &
 \underline{\valci{64.5}{55.0}{73.9}} &
 \valci{66.0}{57.5}{74.5} \\
 & DWI &
 \valci{83.7}{57.3}{110.2} &
 \textit{\valci{75.9}{71.2}{80.6}} &
 \valci{83.7}{76.3}{91.0} &
 \valci{84.9}{77.2}{92.6} &
 \underline{\valci{75.7}{67.9}{83.5}} &
 \textbf{\valci{74.4}{67.1}{81.7}} \\
 & Ensemble &
 \valci{256.0}{0.0}{256.0} &
 \valci{60.2}{25.9}{94.4} &
 \textit{\valci{60.1}{49.7}{70.5}} &
 \textbf{\valci{47.3}{24.2}{70.4}} &
 \valci{67.5}{59.7}{75.3} &
 \underline{\valci{56.3}{48.1}{64.5}} \\
\bottomrule
\end{tabularx}
\label{tab:hd95}
\end{table}

\noindent\textbf{ATLAS:} The ATLAS test set is considered in-domain. The \texttt{baseline} model is expected to perform optimally, while the qMRI generator model must accurately reproduce MPRAGE details to match this performance. As shown in Tables \ref{tab:dice} \& \ref{tab:hd95} and Figure \ref{fig:combined-box}, the \texttt{baseline} outperformed the qMRI-based methods, with no statistically significant differences except between \texttt{Synth} and \texttt{qSynth}. Notably, \texttt{qSynth} significantly outperformed \texttt{Synth}, indicating that the physics-constrained pipeline reduces domain shift between simulated and real data.

\noindent\textbf{ARC:} The ARC dataset exhibits moderate domain shift in the T1w channel and larger shifts in T2 and FLAIR. As shown in Tables \ref{tab:dice} \& \ref{tab:hd95} and Figure \ref{fig:combined-box}, \texttt{qATLAS} underperformed relative to the \texttt{baseline} (though not significantly). For T1w, \texttt{qSynth} initially lagged behind \texttt{Synth} but reached comparable performance when real data was integrated; in T2w, \texttt{qSynth} outperformed all models, and in FLAIR and ensemble evaluations, it achieved statistically significant improvements, highlighting its robustness under domain shift. 

\noindent\textit{Why do real scans help more in some contrasts than others?} 
We currently model the entire lesion with a \emph{single} Gaussian intensity prior.  This approximation is adequate for T2w or FLAIR images, where stroke lesions are nearly isointense within the lesion core, so mixing in real data offers only marginal gain and can even dilute the synthetic coverage.  In T1w scans stroke appearance is markedly more heterogeneous, so a single-mode prior under-represents the true lesion distribution. Injecting real ATLAS images therefore supplies the missing high-variance examples and yields the largest Dice improvement for T1w.  Future work could address this imbalance by using multi-modal lesion priors.

\noindent\textbf{PLORAS:} The PLORAS dataset, containing real clinical data from UK hospitals, highlights the superiority of \texttt{qSynth} over \texttt{baseline} and \texttt{qATLAS} models as seen in Tables \ref{tab:dice} \& \ref{tab:hd95}. \texttt{qSynth} also demonstrated moderate but consistent improvements over \texttt{Synth}, reflecting its capacity to generalise effectively to diverse real-world clinical scenarios.

\noindent\textbf{ISLES 2015:} The ISLES 2015 dataset, featuring co-registered T1w, T2w, FLAIR, and DWI channels with acute stroke lesions, posed unique challenges. DWI channels, absent in qMRI simulations, were unseen during \texttt{qATLAS}/\texttt{qSynth} training. Nevertheless, \texttt{qSynth} significantly outperformed all models, including Synth, which could have trained on DWI-like data. \texttt{qSynth} achieved the highest Dice scores for T2w and FLAIR modalities, while both \texttt{Synth} and \texttt{qSynth} showed strong ensemble performance. For T1w, \texttt{qATLAS} and \texttt{Synth} achieved the best results. The dataset’s skull-stripped images likely conferred an advantage to \texttt{Synth}/\texttt{qSynth} models compared to other methods.

\section{Conclusion}

We proposed two physics-aware generators for stroke-lesion segmentation training: \texttt{qATLAS}, which derives qMRI maps from a single MPRAGE, and \texttt{qSynth}, which samples qMRI from learned intensity priors. \texttt{qATLAS} boosts T1-weighted performance but degrades on contrasts dominated by non-$R_{1}$ parameters; \texttt{qSynth} delivers the most consistent cross-domain gains, yet both remain below fully in-domain models. Future work should explore domain adaptation techniques or domain-conditioning via hypernetworks to further close this gap. Additionally, our framework can be extended to other tasks (e.g., registration, super-resolution), other pathologies (e.g. glioblastoma), and other anatomies (e.g., cardiac, abdomen). We also plan to integrate recent advances in more realistic lesion simulations \cite{liu2025unravelingnormalanatomyfluiddriven} to further enhance synthetic data fidelity. Overall, embedding MRI physics in data generation represents a promising step toward robust, generalisable segmentation in diverse clinical scenarios.

\begin{credits}
\subsubsection{\ackname} LC is supported by the EPSRC-funded UCL Centre for Doctoral Training in Intelligent, Integrated Imaging in Healthcare (i4health) (EP/S021930/1), and the Wellcome Trust (203147/Z/16/Z and 205103/Z/16/Z). CP is funded by Wellcome (203147/Z/16/Z, 205103/Z/16/Z and 224562/Z/21/Z). This research was supported by NVIDIA and used NVIDIA RTX A6000 48GB.

\subsubsection{\discintname}
The authors have no competing interests to declare that are
relevant to the content of this article.
\end{credits}

\bibliography{citations}

\begin{thebibliography}{10}
\providecommand{\url}[1]{\texttt{#1}}
\providecommand{\urlprefix}{URL }
\providecommand{\doi}[1]{https://doi.org/#1}

\bibitem{Billot2021}
Billot, B., Cerri, S., Leemput, K.V., et~al.: Joint segmentation of multiple sclerosis lesions and brain anatomy in {MRI} scans of any contrast and resolution with {CNNs}. In: 2021 IEEE 18th International Symposium on Biomedical Imaging (ISBI). IEEE (Apr 2021). \doi{10.1109/isbi48211.2021.9434127}

\bibitem{Billot2023}
Billot, B., Greve, D.N., Puonti, O., et~al.: {SynthSeg}: Segmentation of brain {MRI} scans of any contrast and resolution without retraining. Medical Image Analysis  \textbf{86},  102789 (May 2023). \doi{10.1016/j.media.2023.102789}

\bibitem{Borges2023}
Borges, P., Fernandez, V., Tudosiu, P.D., et~al.: Unsupervised Heteromodal Physics-Informed Representation of MRI Data: Tackling Data Harmonisation, Imputation and Domain Shift, p. 53–63. Springer Nature Switzerland (2023). \doi{10.1007/978-3-031-44689-4\_6}, \url{http://dx.doi.org/10.1007/978-3-031-44689-4\_6}

\bibitem{borges2021acquisitioninvariant}
Borges, P., Shaw, R., Varsavsky, T., et~al.: Acquisition-invariant brain mri segmentation with informative uncertainties (2021)

\bibitem{Brudfors_2020}
Brudfors, M., Balbastre, Y., Flandin, G., et~al.: Flexible {B}ayesian Modelling for Nonlinear Image Registration, p. 253–263. Springer International Publishing (2020). \doi{10.1007/978-3-030-59716-0\_25}

\bibitem{Brzus2023}
Brzus, M., Boes, A.D., Bruss, J., Johnson, H.J.: Leveraging high-quality research data for ischemic stroke lesion segmentation on clinical data. In: 2023 IEEE 20th International Symposium on Biomedical Imaging (ISBI). p. 1–5. IEEE (Apr 2023). \doi{10.1109/isbi53787.2023.10230775}, \url{http://dx.doi.org/10.1109/ISBI53787.2023.10230775}

\bibitem{cardoso2022monai}
Cardoso, M.J., Li, W., Brown, R., et~al.: Monai: An open-source framework for deep learning in healthcare (2022)

\bibitem{chalcroft2024syntheticdatarobuststroke}
Chalcroft, L., Pappas, I., Price, C.J., Ashburner, J.: Synthetic data for robust stroke segmentation (2024), \url{https://arxiv.org/abs/2404.01946}

\bibitem{chen2019med3d}
Chen, S., Ma, K., Zheng, Y.: Med3d: Transfer learning for 3d medical image analysis (2019)

\bibitem{Dorent2023}
Dorent, R., Kujawa, A., Ivory, M., et~al.: Crossmoda 2021 challenge: Benchmark of cross-modality domain adaptation techniques for vestibular schwannoma and cochlea segmentation. Medical Image Analysis  \textbf{83},  102628 (Jan 2023). \doi{10.1016/j.media.2022.102628}, \url{http://dx.doi.org/10.1016/j.media.2022.102628}

\bibitem{Gibson2024TheRepository}
Gibson, M., Newman-Norlund, R., Bonilha, L., et~al.: {The Aphasia Recovery Cohort, an open-source chronic stroke repository}. Scientific Data  \textbf{11}(1), ~1--8 (2024). \doi{10.1038/s41597-024-03819-7}, \url{http://dx.doi.org/10.1038/s41597-024-03819-7}

\bibitem{he2015delving}
He, K., Zhang, X., Ren, S., Sun, J.: Delving deep into rectifiers: Surpassing human-level performance on {ImageNet} classification (2015)

\bibitem{hendrycks2023gaussian}
Hendrycks, D., Gimpel, K.: Gaussian error linear units (gelus) (2023)

\bibitem{Isensee2020}
Isensee, F., Jaeger, P.F., Kohl, S.A.A., et~al.: {nnU-Net}: a self-configuring method for deep learning-based biomedical image segmentation. Nature Methods  \textbf{18}(2),  203–211 (Dec 2020). \doi{10.1038/s41592-020-01008-z}

\bibitem{Johnson2024ProgressiveStroke}
Johnson, L., Newman-Norlund, R., Teghipco, A., et~al.: {Progressive lesion necrosis is related to increasing aphasia severity in chronic stroke}. NeuroImage: Clinical  \textbf{41} (1 2024). \doi{10.1016/j.nicl.2024.103566}

\bibitem{Liew2022}
Liew, S.L., Lo, B.P., Donnelly, M.R., et~al.: A large, curated, open-source stroke neuroimaging dataset to improve lesion segmentation algorithms. Scientific Data  \textbf{9}(1) (Jun 2022). \doi{10.1038/s41597-022-01401-7}

\bibitem{Liu2021}
Liu, C.F., Hsu, J., Xu, X., et~al.: Deep learning-based detection and segmentation of diffusion abnormalities in acute ischemic stroke. Communications Medicine  \textbf{1}(1) (Dec 2021). \doi{10.1038/s43856-021-00062-8}, \url{http://dx.doi.org/10.1038/s43856-021-00062-8}

\bibitem{liu2025unravelingnormalanatomyfluiddriven}
Liu, P., Aguila, A.L., Iglesias, J.E.: Unraveling normal anatomy via fluid-driven anomaly randomization (2025), \url{https://arxiv.org/abs/2501.13370}

\bibitem{loshchilov2019decoupled}
Loshchilov, I., Hutter, F.: Decoupled weight decay regularization (2019)

\bibitem{Maier2017}
Maier, O., Menze, B.H., von~der Gablentz, J., et~al.: {ISLES} 2015 - a public evaluation benchmark for ischemic stroke lesion segmentation from multispectral {MRI}. Medical Image Analysis  \textbf{35},  250–269 (Jan 2017). \doi{10.1016/j.media.2016.07.009}

\bibitem{delarosa2024robustensemblealgorithmischemic}
de~la Rosa, E., Reyes, M., Liew, S.L., et~al.: A robust ensemble algorithm for ischemic stroke lesion segmentation: Generalizability and clinical utility beyond the isles challenge (2024), \url{https://arxiv.org/abs/2403.19425}

\bibitem{Tabelow2019}
Tabelow, K., Balteau, E., Ashburner, J., et~al.: hmri – a toolbox for quantitative mri in neuroscience and clinical research. NeuroImage  \textbf{194},  191–210 (Jul 2019). \doi{10.1016/j.neuroimage.2019.01.029}, \url{http://dx.doi.org/10.1016/j.neuroimage.2019.01.029}

\bibitem{ulyanov2017instance}
Ulyanov, D., Vedaldi, A., Lempitsky, V.: Instance normalization: The missing ingredient for fast stylization (2017)

\bibitem{varadarajan2021unsupervised}
Varadarajan, D., Bouman, K.L., van~der Kouwe, A., et~al.: Unsupervised learning of mri tissue properties using mri physics models (2021)

\bibitem{wang2021tent}
Wang, D., Shelhamer, E., Liu, S., et~al.: Tent: Fully test-time adaptation by entropy minimization (2021)

\bibitem{Wang_2019}
Wang, G., Li, W., Ourselin, S., Vercauteren, T.: Automatic Brain Tumor Segmentation Using Convolutional Neural Networks with Test-Time Augmentation, p. 61–72. Springer International Publishing (2019). \doi{10.1007/978-3-030-11726-9\_6}, \url{http://dx.doi.org/10.1007/978-3-030-11726-9\_6}

\bibitem{zhang2018unreasonable}
Zhang, R., Isola, P., Efros, A.A., et~al.: The unreasonable effectiveness of deep features as a perceptual metric (2018)

\end{thebibliography}

\end{document}